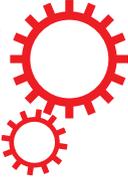

# Palladium gates for reproducible quantum dots in silicon

Matthias Brauns[1,2], Sergey V. Amitonov[1], Paul-Christiaan Spruijtenburg[1] & Floris A. Zwanenburg[1]

We replace the established aluminium gates for the formation of quantum dots in silicon with gates made from palladium. We study the morphology of both aluminium and palladium gates with transmission electron microscopy. The native aluminium oxide is found to be formed all around the aluminium gates, which could lead to the formation of unintentional dots. Therefore, we report on a novel fabrication route that replaces aluminium and its native oxide by palladium with atomic-layer-deposition-grown aluminium oxide. Using this approach, we show the formation of low-disorder gate-defined quantum dots, which are reproducibly fabricated. Furthermore, palladium enables us to further shrink the gate design, allowing us to perform electron transport measurements in the few-electron regime in devices comprising only two gate layers, a major technological advancement. It remains to be seen, whether the introduction of palladium gates can improve the excellent results on electron and nuclear spin qubits defined with an aluminium gate stack.

The realization of a quantum computer using spins in a solid-state system[1,2] has made impressive progress over the last decades. In recent years, group-IV materials like silicon[3] and carbon[4] have attracted a lot of attention, since they can be isotopically purified to only consist of spin-zero nuclei. Metal-oxide-semiconductor devices inspired by classical transistors have proven to be highly suitable for the realization of quantum bits both in intrinsic silicon and silicon-germanium heterostructures[3,5]. Their very flexible design has enabled single and double quantum dots[6–9], spin read-out via Pauli spin blockade[10–15], charge sensing experiments with a quantum point contact[16] and dispersive read-out[17], single qubits[10,18–21] and two-qubit logic gates[22] in quick succession. With the demonstration of these building blocks, the reproducible fabrication of fully gate-tuneable devices receives increased attention[23,24]. The formation of unintentional quantum dots[25–27] poses a substantial problem, since they can capacitively couple to the intended quantum dot and disturb both transport and charge sensing measurements. The choice of the gate material plays a central role here, since e.g. different thermal expansion coefficients within the device lead to mechanical strain leading to fluctuations in the electrochemical potential[27,28]. Also chemical reactions with surrounding dielectric layers have to be taken into account due to the possible formation of charge traps. Finally, also the morphology of the gate material dictates reachable feature sizes and gate design. Due to its high-quality native oxide that renders the deposition of inter-gate dielectric layers unnecessary, aluminium was, in recent years, the most commonly used gate material for accumulation-mode quantum dots in silicon, introduced by Angus et al. in 2007[6]. The device design reported there has been the workhorse for the impressive follow-up experiments performed at the University of New South Wales[18,22,29–34], and was also successfully implemented by other research groups[26,35–40]. Besides aluminium, also poly silicon has been employed as a gate material[9,41]. Noble metals, like palladium[7], were so far only used for depletion-type quantum dots that do not require multi-layer gate stacks. For such depletion-type dots with palladium gates, also the role of mechanical stress induced by the electrode for the formation of unintentional quantum dots has been studied and found to be non-negligible[28]. This situation is hard to compare to our accumulation-mode devices with several gate layers, and our devices show no pronounced signature of stress-induced quantum dots.

In this Report, we propose the use of palladium as a gate material for accumulation gates, and compare its performance to the commonly used aluminium. In a first part, we characterize the suitability of the two materials as nanoscale gates by means of transmission electron microscopy (TEM). Subsequently, we assess the usability of devices fabricated with Pd gates for the formation of electrostatically defined quantum dots.

[1]NanoElectronics Group, MESA+ Institute for Nanotechnology, University of Twente, P.O. Box 217, 7500, AE Enschede, The Netherlands. [2]Present address: Institute of Science and Technology Austria, Am Campus 1, 3400, Klosterneuburg, Austria. Correspondence and requests for materials should be addressed to M.B. (email: matthias.brauns@ist.ac.at) or F.A.Z. (email: f.a.zwanenburg@utwente.nl)

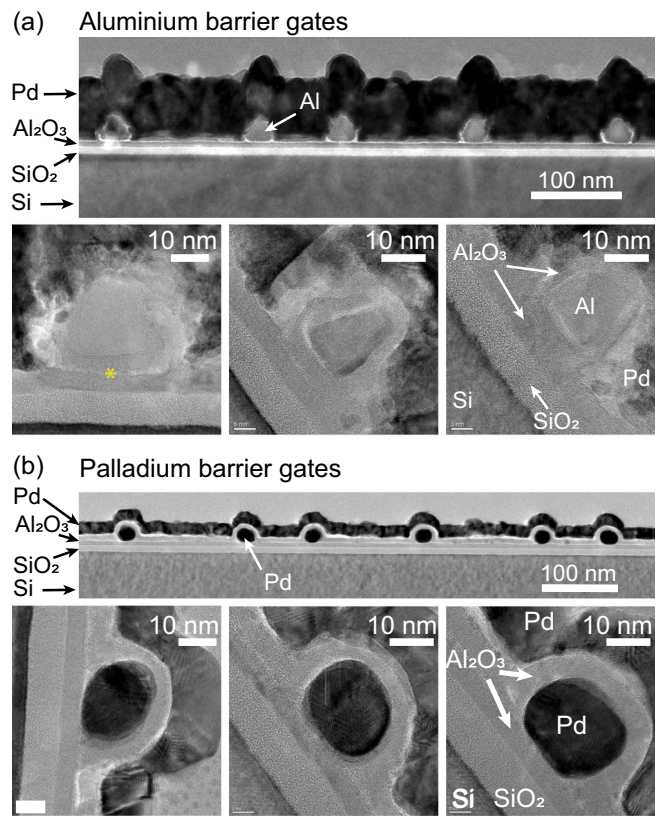

**Figure 1.** Transmission electron microscopy cross sections of (**a**) aluminium and (**b**) palladium gates on top of a Si-SiO$_2$-Al$_2$O$_3$ layer stack capped by Al$_2$O$_3$ and palladium.

## Transmission Electron Microscopy

The structures used for the TEM studies are displayed in Fig. 1. The sample layout follows the pioneering device design of Angus *et al.*[6]. A quasi-intrinsic Si wafer is thermally oxidized to form approximately 8 nm of high-quality silicon oxide. In contrast to the original Angus design, we grow an additional dielectric layer of 5 nm aluminium oxide by atomic layer deposition (ALD) at 250 °C with tetramethylammonium hydroxide and H$_2$O as precursors. In the aluminium gate samples, this layer protects the SiO$_2$ from partially being reduced to SiO$_x$ with $x < 2$. This reduction due to the stronger oxidation affinity of aluminium otherwise leads to defects directly underneath the gates. The gate structures are then formed by patterning a PMMA layer with electron-beam lithography at an acceleration voltage of 28 kV and subsequent electron-beam evaporation of the gate metal followed by lift-off. At least two layers of gates are required for quantum dot definition. We thermally oxidize the aluminium samples at 180 °C under ambient conditions on a hot plate to form a 5 nm layer of Al$_2$O$_3$ around the Al gates in Fig. 1(a). Since palladium is a noble metal, this is not possible in these samples, and we therefore use another ALD step at a reduced temperature of 150 °C to achieve 5 nm of Al$_2$O$_3$ for the samples in Fig. 1(b). For the TEM samples, a thick layer of Pd (thick dark layer in Fig. 1(a)) is deposited on top to protect the sample during preparation of the thin cross section.

The gates in Fig. 1 have a nominal width of 20–30 nm. The effective width of the gate is decreased by approximately 10 nm in the case of Al due to the partial oxidation of the metal. In Fig. 1(a) the different materials are labelled as confirmed by energy-dispersed X-ray (EDX) spectroscopy. In the upper overview panel, we can clearly observe the light SiO$_2$ on top of the Si substrate followed by the grey Al$_2$O$_3$, on top of which the Al gates give a similar contrast to the Al$_2$O$_3$. For more clarity three higher-resolution zooms of single barriers are shown in the lower panel of Fig. 1(a). Two distinct Al$_2$O$_3$ layers are visible, with a very thin interfacial layer (see asterisk in the left single-gate image of Fig. 1(a)) inbetween. Deducing from the fabrication protocol, we identify the two layers as the ALD-grown lower layer, and the thermal-oxidation layer around the gate. Note here that he oxidation process takes place all around the gate, i.e. not only at the aluminium surface directly subjected to air, but also at the interface between the aluminium and the ALD-grown Al$_2$O$_3$ (marked by an asterisk in the lower-left panel of Fig. 1(a)). Here, the oxidation of the aluminium includes the diffusion and subsequent incorporation of oxygen atoms into the material. We speculate that this volumetric increase can generate compressive stress on the underlying layers. The associated band structure modulations can then lead to localization of charges at low temperatures, in analogy to quantum dot formation due to different thermal expansion coefficients[27]. Within the scope of this Report, it is difficult to distinguish these two mechanisms, although we note that the difference in the thermal expansion coefficients with silicon and silicon oxide is very large for aluminium, two times larger than for palladium, while the difference is almost zero for poly-silicon[27,42]. This leads to a reduced risk of charge localization by stress-induced potential fluctuations.

Lim et al.[8] observed a similar aluminium oxide formation at the Al-SiO$_2$ interface in their devices. Quantum dots formed below a single gate have been successfully used to demonstrate single-hole tunneling[37,43] and electron quantum bits[18]. For individual control over the electrochemical potential of the quantum dot as well as the tunnel barriers, however, these strain-induced quantum dots are detrimental.

The Pd gates in Fig. 1(b) do not oxidize, so the second Al$_2$O$_3$ layer grown by ALD is here only visible at the top part of the Pd gates, not all around them. The zoom-in micrographs in the lower panel of Fig. 1(b) exhibit a more uniform shape and width-to-height ratio than their Al counterparts. We can assign this to an interesting difference between the Al and Pd films deposited for lift-off in the same electron-beam evaporator under the same conditions: while the Pd film morphology looks identical on the substrate and on top of the PMMA layer, the Al grains are substantially larger on top of the PMMA layer (see Supplementary Figure 1), leading to the characteristic triangular cross-sections for narrow Al gates. The increased grain size on top of the polymer resist layer suggests a substantial surface diffusion of Al. This limits the structure size achievable with Al compared to Pd under otherwise identical conditions and leads to overall more uniform gate structures for Pd gates in our metal evaporation system. Apart from the grain size, also the Al gate oxidation itself limits the reduction of the device feature size, since 10 nm of the oxide have to be taken into account in the quantum dot design. Center-to-center gate pitches of 40 nm or less are thus hard to reach with aluminium in a lift-off process, but achievable with palladium, and desirable for reaching the few-electron regime (see e.g. the following section).

In conclusion, the TEM study suggests that gate structures similar to the Angus design made from palladium instead of aluminium can achieve smaller feature sizes and fewer unintentional dots in the formation of quantum dots with individual control over the tunnel barriers and dot potential by means of several gates.

### Reproducible quantum dot devices

In Fig. 2(a), a typical device used for studying the fabrication reproducibility of tuneable quantum dot devices made with Pd gates is displayed. 25 nm wide Pd barrier gates with a center-to-center distance of 60 nm are covered by a 5 nm thick Al$_2$O$_3$ dielectric layer, followed by a lead gate also made of Pd with a width of 30 nm between the barrier gates. An additional 5 nm Al$_2$O$_3$ is grown on top of the whole gate structure before the devices are annealed for 30 min at 400 C in a hydrogen atmosphere. The quantum dot formed under the lead gate between the barrier gates thus has a size of approximately 35 by 30 nm.

In Fig. 2(b) we show data of palladium devices from three separate chips, i.e. apart from the SiO$_2$ and lowest Al$_2$O$_3$ layer, they have been fabricated in three separate fabrication runs in order to demonstrate chip-to-chip reproducibility of the current-voltage measurements. For comparison we also added data from a device with Al gates. All measurements shown in Fig. 2 have been performed at 4.2 K. A constant bias voltage of $V_{SD} = 1$ mV was applied to the ohmic contact overlapping with the lead gate on the left as indicated in the sketch in Fig. 2(a), and the current flowing from the drain to ground was measured. The left column in Fig. 2(b) contains current versus gate voltage curves for all three devices. We will call the green curves in Fig. 2(b), where the voltage is applied to the lead gate and both barrier gates simultaneously, turn-on curves. The measurements for the blue (orange) curves in the palladium devices were performed by applying a constant voltage of 4 V to the lead gate and the right (left) barrier gate to ensure electron accumulation between them, and changing the voltage on the left (right) barrier from 4 V to 0 V. We will refer to them as pinch-off curves for the left (right) barrier gate. Measurements on the aluminium device were performed in the same way, except that the constant voltage on the gates not used for pinch-off was 3.5 V.

While all Pd devices turn on at voltages between 2 and 3 V, the pinch-off curves show a significantly lower and much steeper threshold voltage for pinching off the conductance channel, very much like Angus et al.[6] observed. In all three devices the pinch-off curves exhibit only very few or even no resonance around the threshold voltage. Such resonances are commonly attributed to resonant tunneling via localized states within the created tunnel barrier, as discussed in the previous section. We first compare these pinch-off curves to those measured on Al gate devices fabricated in the same cleanroom. Data of one such device is plotted in the lowest panel of Fig. 2(b). Mueller et al. recently reported on a second device, with pinch-off curves in Fig. 2a of their publication[38]. For both Al devices, the pinch-off curves are taken at a $V_{SD} = 1$ mV (same as for the Pd devices), and show multiple resonances each. Multiple kinks and resonances are also visible in the high-bias pinch-off curves in Fig. 2b of Mueller et al.[38]. None of the Al barrier gates displays a clean, resonance-free pinch-off as is the case for barrier 2 of Device C in this Report. We acknowledge the limited statistics of this comparison, but also note that the Al devices reported here and in Mueller's manuscript already represent a significant improvement compared to another report by the same authors[26] (see Fig. 3a there). Data similar to what Mueller et al. reported have also been published by Betz et al.[36], where turn-on as well as pinch-off curves exhibit multiple, irregular resonances, indicating the formation of more than one unintentional quantum dot.

A clearer picture of the device physics can be drawn based on the data shown next to the respective current vs. gate voltage curves in Fig. 2(b). Here, the voltages on the barriers $V_{b1}$ and $V_{b2}$ are varied while a constant voltage $V_{lead} = 4$ V is applied to the lead gate. All three charge stability diagrams of the Pd devices reveal diagonal lines, indicating the formation of a single quantum dot with equal capacitive coupling to both barriers. For Device A, a single instability around $V_{b1} = 1780$ mV is visible. Device B also mainly exhibits one resonance below one of the barriers at $V_{b2} = 1320$ mV, which leads to deviations from the ideal single quantum dot picture indicating the formation of a strongly coupled double quantum dot consisting of the quantum dot between the barriers and one unintentional dot below barrier 2[44]. Finally, Device C is completely free of deviations from the ideal single quantum dot behaviour across many charge transitions, suggesting a defect-free electrostatic environment of the quantum dot. Again, we compare these data to the lowest panel of Fig. 2(b) and those reported by Mueller et al. in Fig. 4b of the publication[38]. In both cases, the data show diagonal Coulomb peaks as a clear indication of the formation of a quantum dot defined by both barrier gates. Additionally, multiple resonances capacitively coupled to only one barrier are visible for both barriers, an indication for more than one unintentional dot. Furthermore, the diagonal Coulomb peaks display 'switchy' behavior as a result of charge noise.

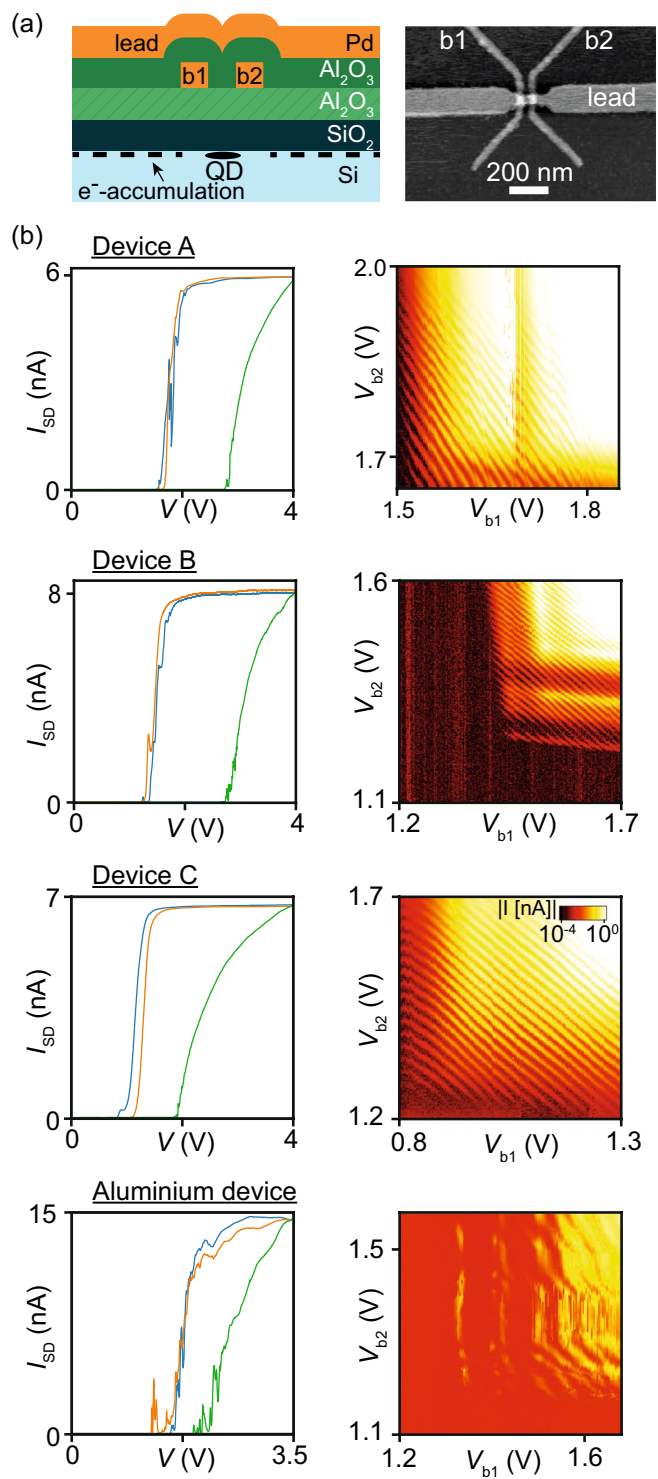

**Figure 2.** (**a**) Schematic cross-section and atomic-force microscopy top-view of a typical Pd device. (**b**) Measurements taken at 4.2 K for three Pd devices on three different chips, and one device with aluminium gates. Left column: current vs gate voltage applied to all three gates simultaneously (green curve), and applied to barrier gate b1 (b2) [blue (orange) curve] while keeping the voltage on the other two gates at 4 V for the Pd devices, and 3.5 V for the Al device. Right column: current versus voltages on b1 and b2 with $V_{lead} = 4\,V$ for the Pd devices, and $V_{lead} = 3.5\,V$ for the Al device.

We want to stress one thought: while comparing data for Pd and Al devices fabricated in the same cleanroom minimizes the effect of using different equipment (e.g. contamination of the evaporation chamber), it also limits the generalizability of the conclusions drawn. We acknowledge that, while the Pd devices seem to exhibit fewer unintentional dots than the Al devices, these findings are not hard proof yet, but more an intriguing hint, that hopefully encourages other research groups to confirm our results.

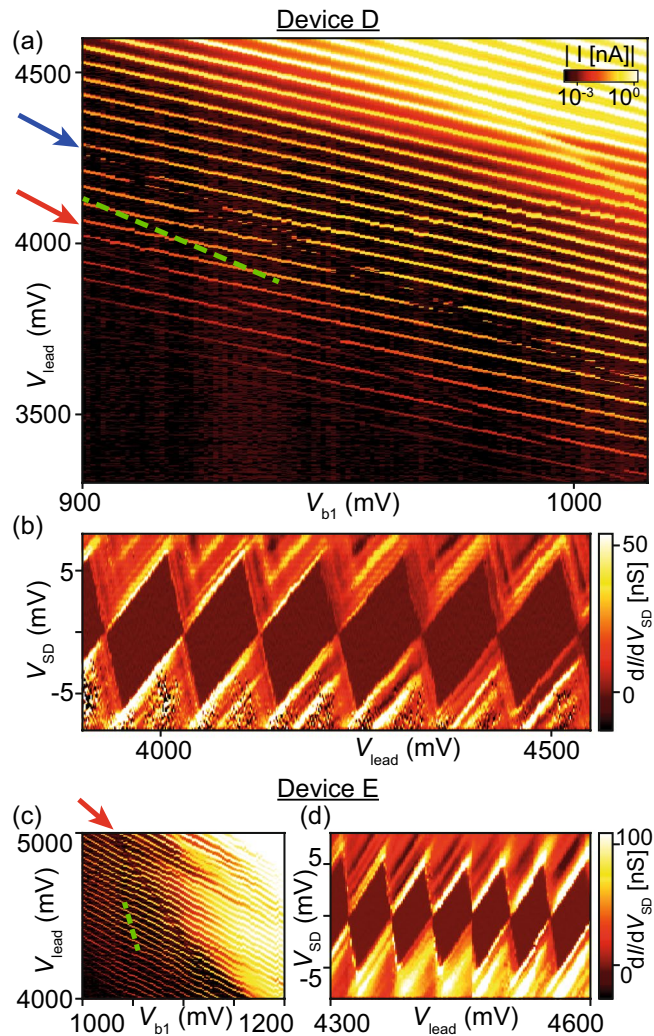

**Figure 3.** Measurements performed on Pd devices at 25 mK. Current versus voltage on the lead gate and the barrier gates at $V_{SD} = 1$ mV in Device D (**a**) and Device E (**c**). In (**a**), $V_{b1}$ and $V_{b2}$ are linked by $V_{b2} = V_{b1} + 300$ mV. In (**c**), $V_{b2} = V_{b1} - 250$ mV, and the colour scale is the same as in (**a**). The arrows indicate deviations from the single quantum dot picture. (**b**) and (**d**) are bias spectroscopy plots along the dashed lines in (**a**) and (**c**), respectively.

Further evidence for the findings from the barrier 1 versus barrier 2 plots is provided by measuring Devices D and E from the same chip as Device A in a dilution refrigerator with an electron temperature of approximately 25 mK. In Fig. 3(a) and (c) we plot charge stability diagrams, where we change $V_{lead}$ versus $V_{b1}$ and $V_{b2}$, with a fixed offset of $V_{b2} = V_{b1} + 300$ mV. Again, parallel, equally spaced current peaks indicate the formation of a single quantum dot. These Coulomb oscillations are only disturbed along two parallel lines marked by a red and a blue arrow in (a), and a single line marked by a red arrow in (c). These disturbances are most likely caused by a defect capacitively coupled to our quantum dot, varying the electrostatic environment by changing its own charge state. Since these two charging events follow parallel lines in gate space in Fig. 3(a), they most likely have the same origin.

Bias spectroscopy measurements are displayed for both devices in Fig. 3(b) and (d), where we plot the numerical differential conductance $dI/dV_{SD}$ while changing $V_{SD}$ and the gate voltages on all three gates simultaneously to follow the dashed lines in Fig. 3(a) and (c). In both cases, the measurements reveal Coulomb diamonds of constant height and shape, indicating a stable electrostatic environment and quantum dot shape and size. The charging energy is approximately 6 meV in both cases. The additonal lines of increased conductance at finite bias can be explained by, e.g., orbital excited states[45,46].

In conclusion, our experiments show that, by using palladium gates, electrostatic definition of quantum dots can be reproducibly achieved and the data suggest a possible reduction of unintentional quantum dots under single barriers compared to Al gate devices. Further evidence under different fabrications conditions is needed to make a general statement.

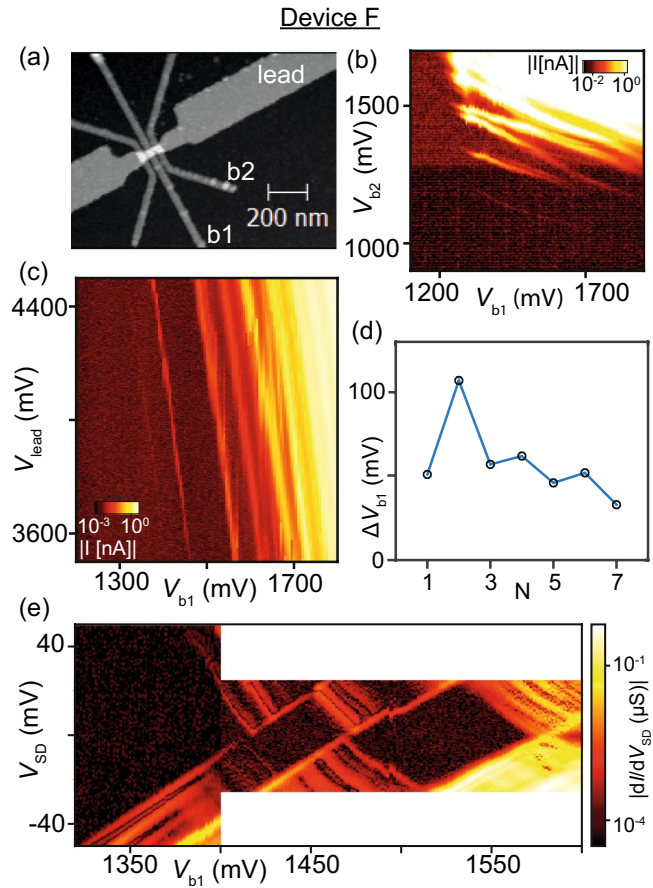

**Figure 4.** (**a**) AFM image of Device F with a gate pitch of 40 nm. The leftmost barrier gate was not connected. (**b**) plot of $I$ versus the voltages on the barrier gates b1 and b2 with fixed $V_{lead} = 4000$ mV and $V_{SD} = 1$ mV. (**c**) $I$ plotted versus $V_{lead} = 4000$ mV and $V_{b1} = V_{b2} + 300$ mV at $V_{SD} = 1$ mV. (**d**) The Coulomb peak distance $\Delta V_{b1}$ between adjacent Coulomb peaks at $V_{lead} = 3500$ mV. (**e**) Differential conductance $dI/dV_{SD}$ versus $V_{b1}$ at $V_{lead} = 3500$ mV. All measurements taken at 4.2 K.

### Few-electron quantum dots with two-layer devices

Reaching the few-electron regime in gate-defined quantum dots in silicon has proven to be very difficult due to the comparably high effective mass of the electrons[3], a problem that has successfully been circumvented by employing a more complex gate stack with a third layer comprising a dedicated plunger gate[47].

Palladium as a gate metal provides us with an opportunity to reach the few-electron regime without complicating the device design: shrinking the size of the quantum dot to hold fewer electrons to start with. Since they do not oxidize, we can shrink the width of the barrier gates as well as their thickness to approximately 10–15 nm without risking broken gates due to oxidation. This allows us to scale down the barrier gate pitch to 40 nm in Device F (Fig. 4(a)). The charge stability diagram in Fig. 4(b), where we change $V_{b1}$ and $V_{b2}$, displays again diagonal Coulomb oscillation, but far fewer than in the 60 nm pitch devices in Fig. 3.

A further signature for the lower number of electrons on this quantum dot is revealed in Fig. 4(c), where we plot $I_{SD}$ versus $V_{lead}$ and $V_{b1}$, with $V_{b2} = V_{b1} - 300$ mV. The Coulomb peaks are parallel, signature for the formation of a single quantum dot, but not equidistant. The peak spacing $\Delta V_{b1}$ in Fig. 4(d) shows a clear even-odd effect, as well as a trend towards lower peak spacings for higher numbers of electrons $N$ added to the quantum dot. The latter indicates an increasing gate capacitance and thus a deviation from the constant interaction model[48]. This can be explained by a decreasing quantum dot size with decreasing electron occupation[49]. The current is non-zero between Coulomb peaks for $V_{b1} \gtrsim 1550$ mV at $V_{lead} = 4000$ mV, a sign for the tunnel barriers being transparent enough for inelastic processes to become significant[46]. At lower barrier gate voltages, the Coulomb peaks are well-separated by Coulomb blockade, and at $V_{b1} < 1410$ mV for $V_{lead} = 3500$ mV, no Coulomb peak is visible any more. There are two possible explanations for this: either the quantum dot is empty at this point, or the tunnel barriers simply have become too opaque to allow for a measurable Coulomb peak even though there are still electron states available.

Since our device does not feature a charge sensor[16], we are restricted to transport measurements, which makes it difficult to distinguish the two scenarios. A signature of a not yet empty quantum dot is a saw-tooth-like pattern at finite bias[43], whereas a clean opening of the Coulomb diamond up to high source-drain bias voltages[47,50–52] is a strong sign for the absence of available charge states, i.e. an empty quantum dot. Figure 4(e) shows a bias spectroscopy of the first two charge transitions visible in Fig. 4(c). For barrier gate voltages lower than $V_{b1} = 1410$ mV, the onset of conductance opens up without any disturbance up to the measurement range of $V_{SD} = \pm 50$ mV, just as expected for an empty dot. Further evidence for reaching the single-electron regime is provided by the addition energies $E_{add}$ of the first electrons added to the quantum dot: while $E_{add} = 17$ meV for the second electron entering the dot, the energy needed to add the third electron is significantly higher ($E_{add} = 24$ meV). This can be explained by a significant orbital energy $E_{orb}$ leading to an even-odd effect for the charging energies[46]. $E_{orb} \approx 7$ meV is consistent with a line of increased conductance visible for positive $V_{SD}$ in the first Coulomb diamond, which could thus represent the first orbital excited state[45]. The even-odd effect due to the additional orbital energy in Fig. 4(d) and (e) becomes weaker for higher numbers of electrons, in accordance with findings in other systems[51,53–55].

The undisturbed high-bias opening of the last Coulomb diamond combined with the even-odd filling of the first electrons onto the quantum dot provides strong evidence for the observation of the single-electron regime in our device with the 40-nm gate pitch.

### Conclusion
In conclusion, we have demonstrated the suitability of palladium with ALD-grown aluminium oxide for gate stacks in MOSFET-like quantum dot devices in silicon. They provide very good run-to-run reproducibility for the fabrication of single-quantum dot devices with only few defects, possibly since chemical alterations of the silicon oxide layer below the gate are avoided, and mechanical stress imposed by the gate is minimized. The small grain size and nobility of palladium also allow for device dimensions small enough to reach the few-electron regime even in two-layer gate designs, something very unusual in accumulation-mode devices in silicon. Performing transport measurements through quantum dots down to the last electron is highly desirable in fundamental research, since a direct energy scale is provided by the applied source-drain bias, which makes spectroscopy measurements much more feasible than by using charge sensing techniques. It also facilitates the read-out of single-spin quantum bits via Pauli spin blockade.

**Data availability.** The datasets generated during and/or analysed during the current study are available from the corresponding author on reasonable request.

### Acknowledgements

The authors want to thank Joost Ridderbos and Wilfred van der Wiel for fruitful discussions. Tess Meijerink was so kind as to be of excellent support during the electron transport measurements. All authors want to express their gratitude to Hans Mertens and Rico Keim for technical support. This work is part of the research programme "Atomic physics in the solid state" with project number 14167, which is (partly) financed by the Netherlands Organisation for Scientific Research (NWO).

### Author Contributions

M.B., S.V.A. and P.C.S. fabricated the devices. M.B. and P.C.S. performed the measurements. M.B., P.C.S. and F.A.Z. analysed the results. M.B. wrote the manuscript. All authors reviewed the manuscript.

### Additional Information

**Supplementary information** accompanies this paper at https://doi.org/10.1038/s41598-018-24004-y.

**Competing Interests:** The authors declare no competing interests.

**Publisher's note:** Springer Nature remains neutral with regard to jurisdictional claims in published maps and institutional affiliations.